\documentclass[aps,superscriptaddress]{revtex4-2}
\usepackage{times}
\usepackage{enumitem,amssymb}
\usepackage{graphicx}
\usepackage{comment}
\usepackage[usenames]{xcolor} 
\usepackage{fancyhdr}

\usepackage{slashed}
\usepackage{graphicx,color}
\usepackage{epsfig}
\usepackage{dcolumn}
\usepackage{bm}
\usepackage{color}
\usepackage{siunitx}

\usepackage{hyperref}
\hypersetup{
    colorlinks=true,
    linkcolor=blue,
    filecolor=magenta,      
    urlcolor=cyan,
    }

\begin{document}

%



\begin{titlepage}

\title{A Call to Arms Control: Synergies between Nonproliferation Applications of Neutrino Detectors and Large-Scale Fundamental Neutrino Physics Experiments \\A Snowmass White Paper}


\medskip




\newcommand{\Editor}{Editor.}
\newcommand{\Contributor}{Primary Contributor.}

\author{Adam Bernstein}
\altaffiliation{\Editor}
\affiliation{\llnl}

\author{Tomi Akindele}
\altaffiliation{\Contributor}
\affiliation{\llnl}

\author{Zelimir Djurcic}
\altaffiliation{\Contributor}
\affiliation{\anl}

\author{Logan Lebanowski}
\altaffiliation{\Contributor}
\affiliation{\penn}

\author{Jost Migenda}
\altaffiliation{\Contributor}
\affiliation{\kings}

\author{Anthony Papatyi}
\altaffiliation{\Contributor}
\affiliation{\pnnl}

\author{Mayly Sanchez}
\altaffiliation{\Contributor}
\affiliation{\iowa}

\author{Michael Smy}
\altaffiliation{\Contributor}
\affiliation{\uci}

\author{Mark Vagins}
\altaffiliation{\Contributor}
\affiliation{\uci}

\collaboration{Editor and Contributors}

\author{Z. Djurcic}\affiliation{\anl}
\author{C. Mariani}	\affiliation{\vtech}
\author{A. Erlandson}\affiliation{Canadian Nuclear Laboratories Ltd., Chalk River, ON, Canada}
\author{R. Mahapatra}\affiliation{Texas A\&M University, College Station, Texas}
\author{M. V. Garzelli}\affiliation{University of Hamburg, Hamburg, Germany}	
\author{E. Kemp}\affiliation{Universidade de Campinas - UNICAMP, Cidade Universitaria "Zeferino Vaz"
CEP 13083-970 - Campinas-SP}
\author{R. Bruce Vogelaar}\affiliation{\vtech}
\author{J. Pedro Ochoa-Ricoux}\affiliation{\uci}
\author{G. Pronost}\affiliation{Kamioka Observatory, ICRR, University of Tokyo}	
\author{S. Seo}\affiliation{IBS Center for Underground Physics, 34126, IBS, 55, Expo-ro, Yuseong-gu, Daejeon,Republic of Korea}
\author{F. Suekane}\affiliation{Research Center for Neutrino Science (RCNS), Tohoku University, Tohuku, Japan}
\author{W. Wang}\affiliation{Sun Yat-sen University, Guangzhou, China}
\author{Y. Keung Hor}\affiliation{Sun Yat-sen University, Guangzhou, China}
\author{M. Chen}\affiliation{Queen's University, Kingston, ON K7L 3N6, Canada}

\newcommand{\anl}{Argonne National Laboratory, Lemont, IL 60439, USA}
\newcommand{\vtech}{Center for Neutrino Physics, Virginia~Tech, Blacksburg, Virginia 24061 USA}

\collaboration{Co-Signers from DUNE, SNO+, JUNO, Super-K,Hyper-K, and other collaborations}
\author{T.~Anderson}\affiliation{\psu}
\author{E.~Anderssen}\affiliation{\lbnl}
\author{M.~Askins}\affiliation{\lbnl}\affiliation{\ucb}
\author{A.~J.~Bacon}\affiliation{\penn}
\author{Z.~Bagdasarian}\affiliation{\lbnl}\affiliation{\ucb}
\author{A.~Baldoni}\affiliation{\psu}
\author{N.~Barros}\affiliation{\fcul}\affiliation{ \lip}
\author{L.~Bartoszek}\affiliation{\bart}
\author{A.~Bat}\affiliation{\erc}
\author{M.~Bergevin}\affiliation{\llnl}
\author{A.~Bernstein}\affiliation{\llnl}
\author{E.~Blucher}\affiliation{\chic}
\author{J.~Boissevain}\affiliation{\bart}
\author{R.~Bonventre}\affiliation{\lbnl}
\author{D.~Brown}\affiliation{\lbnl}
\author{E.~J.~Callaghan}\affiliation{\lbnl}\affiliation{\ucb}
\author{D.~F.~Cowen}\affiliation{\psu}
\author{S.~Dazeley}\affiliation{\llnl}
\author{M.~Diwan}\affiliation{\bnl}
\author{K.~Frankiewicz}\affiliation{\bu}
\author{C.~Grant}\affiliation{\bu}
\author{T.~Kaptanoglu}\affiliation{\lbnl}\affiliation{\ucb}
\author{J.R.~Klein}\affiliation{\penn}
\author{C.~Kraus}\affiliation{\snolab}\affiliation{\laur}
\author{T.~Kroupa}\affiliation{\penn}
\author{B.~Land}\affiliation{\lbnl}\affiliation{\ucb}\affiliation{\penn}
\author{L.~Lebanowski}\affiliation{\penn}
\author{V.~Lozza}\affiliation{\fcul}\affiliation{ \lip}
\author{A.~Marino}\affiliation{\boul}
\author{A.~Mastbaum}\affiliation{\rut}
\author{C.~Mauger}\affiliation{\penn}
\author{S.~Naugle}\affiliation{\penn}
\author{M.~Newcomer}\affiliation{\penn}
\author{A.~Nikolica}\affiliation{\penn}
\author{G.~D.~Orebi~Gann}\affiliation{\lbnl}\affiliation{\ucb}
\author{L.~Pickard}\affiliation{\ucd}
\author{J.~Saba}\affiliation{\lbnl}
\author{S.~Schoppmann}\affiliation{\lbnl}\affiliation{\ucb}
\author{J.~Sensenig}\affiliation{\penn}
\author{M.~Smiley}\affiliation{\lbnl}\affiliation{\ucb}
\author{H.~Steiger}\affiliation{\mainz}\affiliation{\mun}
\author{R.~Svoboda}\affiliation{\ucd}
\author{E.~Tiras}\affiliation{\erc}\affiliation{\iow}
\author{W.~H.~Trzaska}\affiliation{\jyv}
\author{R.~van~Berg}\affiliation{\penn}\affiliation{\bart}
\author{G.~Wendel}\affiliation{\psu}
\author{M.~Wetstein}\affiliation{\iowa}
\author{M.~Wurm}\affiliation{\mainz}
\author{G.~Yang}\affiliation{\lbnl}\affiliation{\ucb}
\author{M.~Yeh}\affiliation{\bnl}
\author{E.D.~Zimmerman}\affiliation{\boul}

\newcommand{\ucb}{Physics Department, University of California at Berkeley, Berkeley, CA 94720-7300
}
\newcommand{\lbnl}{
Lawrence Berkeley National Laboratory, 1 Cyclotron Road, Berkeley, CA 94720-8153, USA
}
\newcommand{\penn}{Department of Physics and Astronomy, University of Pennsylvania, Philadelphia, PA 19104-6396
}\newcommand{\fcul}{Universidade de Lisboa, Faculdade de Ci{\^e}ncias (FCUL), Departamento de F{\'i}sica, Campo Grande, Edifício C8, 1749-016 Lisboa, Portugal
}\newcommand{\lip}{Laborat{\'o}rio de Instrumenta{}{\c c}{\~a}o e F{\'i}sica Experimental de Part{\'i}culas (LIP), Av. Prof. Gama Pinto, 2, 1649-003, Lisboa, Portugal
}\newcommand{\chic}{
The Enrico Fermi Institute and Department of Physics, The University of Chicago, Chicago, IL 60637, USA
}\newcommand{\bnl}{
Brookhaven National Laboratory, Upton, New York 11973, USA
}\newcommand{\uh}{
University of Hawai‘i at Manoa, Honolulu, Hawai‘i 96822, USA
}\newcommand{\iowa}{
Department of Physics and Astronomy, Iowa State University, Ames, IA 50011, USA
}\newcommand{\jyv}{
Department of Physics, University of Jyv{\"a}skyl{\"a}, Finland
}\newcommand{\ucd}{
University of California, Davis, 1 Shields Avenue, Davis, CA 95616, USA
}\newcommand{\bu}{
Boston University, Department of Physics, Boston, MA 02215, USA
}\newcommand{\mainz}{
Institute of Physics and Excellence Cluster PRISMA, Johannes Gutenberg-Universit{\"a}t Mainz, 55099 Mainz, Germany
}\newcommand{\ham}{
Institut f{\"u}r Experimentalphysik, Universit{\"a}t Hamburg, 22761 Hamburg, Germany
}\newcommand{\alb}{
University of Alberta, Department of Physics, 4-181 CCIS, Edmonton, AB T6G 2E1, Canada
}\newcommand{\pnnl}{
Pacific Northwest National Laboratory, Richland, WA 99352, USA
}\newcommand{\laur}{
Laurentian University, Department of Physics, 935 Ramsey Lake Road, Sudbury, ON P3E 2C6, Canada
}\newcommand{\lsu}{
Department of Physics and Astronomy, Louisiana State University, Baton Rouge, LA 70803
}\newcommand{\tub}{
Kepler Center for Astro and Particle Physics, Universit{\"a}t T{\"u}bingen, 72076 T{\"u}bingen, Germany
}\newcommand{\sheff}{
University of Sheffield, Physics \& Astronomy, Western Bank, Sheffield S10 2TN, UK
}\newcommand{\qu}{
Queen's University, Department of Physics, Engineering Physics \& Astronomy, Kingston, ON K7L 3N6, Canada
}\newcommand{\snolab}{
SNOLAB, Creighton Mine 9, 1039 Regional Road 24, Sudbury, ON P3Y 1N2, Canada
}\newcommand{\rut}{
Department of Physics and Astronomy, Rutgers, The State University of New Jersey, 136 Frelinghuysen Road, Piscataway, NJ 08854-8019 USA
}\newcommand{\temp}{
Department of Physics, Temple University, Philadelphia, PA, USA
}\newcommand{\ucla}{
University of California Los Angeles, Department of Physics \& Astronomy, 475 Portola Plaza, Los Angeles, CA 90095-1547, USA
}
\newcommand{\tri}{
SISSA/INFN, Via Bonomea 265, I-34136 Trieste, Italy
}\newcommand{\kav}{
Kavli IPMU (WPI), University of Tokyo, 5-1-5 Kashiwanoha, 277-8583 Kashiwa, Japan
}\newcommand{\kor}{
Center for Underground Physics, Institute for Basic Science, Daejeon 34126, Korea
}\newcommand{\uci}{
University of California, Irvine, Department of Physics, CA 92697, Irvine, USA
}\newcommand{\sbu}{
State University of New York at Stony Brook, Department of Physics and Astronomy, Stony Brook, New York, USA
}\newcommand{\tsing}{
Department of Engineering Physics, Tsinghua University, Beijing 100084, China
}\newcommand{\corn}{
Cornell University, Ithaca, NY, USA
}\newcommand{\boul}{
University of Colorado at Boulder, Department of Physics, Boulder, Colorado, USA
}\newcommand{\dres}{
Institut f{\"u}r Kern und Teilchenphysik, TU Dresden, Zellescher Weg 19, 01069, Dresden, Germany
}
\newcommand{\mun}{Physics Department, Technische Universit{\"a}t M{\"u}nchen, 85748 Garching, Germany
}
\newcommand{\mitnew}{
Massachusetts Institute of Technology, Department of Physics and Laboratory for Nuclear Science, 77 Massachusetts Ave Cambridge, MA 02139, USA
}
\newcommand{\kings}{King’s College London, Department of Physics, Strand Building, Strand, London WC2R 2LS, UK}
\newcommand{\llnl}{
Lawrence Livermore National Laboratory, Livermore, CA 94550, USA
}
\newcommand{\fnal}{
Fermi National Accelerator Laboratory, Batavia, IL 60510, USA
}
\newcommand{\erc}{Department of Physics, Erciyes University, 38030, Kayseri, Turkey
}
\newcommand{\iow}{Department of Physics and Astronomy, The University of Iowa, Iowa City, Iowa, USA}
\newcommand{\psu}{Pennsylvania State University, University Park, PA 16802, USA}

\newcommand{\heid}{Ruprecht-Karls-Universitat Heidelberg, Im
Neuenheimer Feld 227, Heidelberg, Germany}

\newcommand{\bart}{Bartoszek Engineering, Aurora, IL 60506, USA}

\newcommand{\ucbne}{Nuclear Engineering Department, University of California at Berkeley, Berkeley, CA 94720-7300
}
\collaboration{Co-Signers from Eos }
\author{M.~Askins}\affiliation{\lbnl}\affiliation{\ucb}
\author{Z.~Bagdasarian}\affiliation{\lbnl}\affiliation{\ucb}
\author{N.~Barros}\affiliation{\penn}
\affiliation{\fcul}
\affiliation{ \lip}
\author{E.W.~Beier}\affiliation{\penn}
\author{A.~Bernstein}\affiliation{\llnl}
\author{M.~B\"ohles}\affiliation{\mainz}
\author{E.~Blucher}\affiliation{\chic}
\author{R.~Bonventre}\affiliation{\lbnl}
\author{E.~Bourret}\affiliation{\lbnl}
\author{E.~J.~Callaghan}\affiliation{\ucb}\affiliation{\lbnl}
\author{J.~Caravaca}\affiliation{\ucb}\affiliation{\lbnl}
\author{M.~Diwan}\affiliation{\bnl}
\author{S.T.~Dye}\affiliation{\uh}
\author{J.~Eisch}\affiliation{\fnal}
\author{A.~Elagin}\affiliation{\chic}
\author{T.~Enqvist}\affiliation{\jyv}
\author{U.~Fahrendholz}\affiliation{\mun}
\author{V.~Fischer}\affiliation{\ucd}
\author{K.~Frankiewicz}\affiliation{\bu}
\author{C.~Grant}\affiliation{\bu}
\author{D.~Guffanti}\affiliation{\mainz}
\author{C.~Hagner}\affiliation{\ham}
\author{A.~Hallin}\affiliation{\alb}
\author{C.~M.~Jackson}\affiliation{\pnnl}
\author{R.~Jiang}\affiliation{\chic}
\author{T.~Kaptanoglu}\affiliation{\ucb}\affiliation{\lbnl}
\author{J.R.~Klein}\affiliation{\penn}
\author{Yu.~G.~Kolomensky}\affiliation{\ucb}\affiliation{\lbnl}
\author{C.~Kraus}\affiliation{\snolab}\affiliation{\laur}
\author{F.~Krennrich}\affiliation{\iowa}
\author{T.~Kutter}\affiliation{\lsu}
\author{T.~Lachenmaier}\affiliation{\tub}
\author{B.~Land}\affiliation{\ucb}\affiliation{\lbnl}\affiliation{\penn}
\author{K.~Lande}\affiliation{\penn}
\author{L.~Lebanowski}\affiliation{\penn}
\author{J.G.~Learned}\affiliation{\uh}
\author{V.A.~Li}\affiliation{\llnl}
\author{V.~Lozza}\affiliation{\fcul}
\affiliation{ \lip}
\author{L.~Ludhova}\affiliation{\jul}\affiliation{\aach}
\author{M.~Malek}\affiliation{\sheff}
\author{S.~Manecki}\affiliation{\laur}\affiliation{\qu}\affiliation{\snolab}
\author{J.~Maneira}\affiliation{\fcul}
\affiliation{ \lip}
\author{J.~Maricic}\affiliation{\uh}
\author{J.~Martyn}\affiliation{\mainz}
\author{A.~Mastbaum}\affiliation{\rut}
\author{C.~Mauger}\affiliation{\penn}
\author{M.~Mayer}\affiliation{\mun}
\author{J.~Migenda}\affiliation{\kings}
\author{F.~Moretti}\affiliation{\lbnl}
\author{J.~Napolitano}\affiliation{\temp}
\author{B.~Naranjo}\affiliation{\ucla}
\author{S.~Naugle}\affiliation{\penn}
\author{M.~Nieslony}\affiliation{\mainz}
\author{L.~Oberauer}\affiliation{\mun}
\author{G.~D.~Orebi~Gann}\affiliation{\ucb}\affiliation{\lbnl}
\author{J.~Ouellet}\affiliation{\mitnew}
\author{T.~Pershing}\affiliation{\ucd}
\author{S.T.~Petcov}\affiliation{\tri,\kav}
\author{L.~Pickard}\affiliation{\ucd}
\author{R.~Rosero}\affiliation{\bnl}
\author{M.~C.~Sanchez}\affiliation{\iowa}
\author{J.~Sawatzki}\affiliation{\mun}
\author{S.~Schoppmann}\affiliation{\ucb}\affiliation{\lbnl}
\author{S.H.~Seo}\affiliation{\kor}
\author{M.~Smiley}\affiliation{\ucb}\affiliation{\lbnl}
\author{M.~Smy}\affiliation{\uci}
\author{A.~Stahl}\affiliation{\aach}
\author{H.~Steiger}\affiliation{\mainz}\affiliation{\mun}
\author{M.~R.~Stock}\affiliation{\mun}
\author{H.~Sunej}\affiliation{\bnl}
\author{R.~Svoboda}\affiliation{\ucd}
\author{E.~Tiras}\affiliation{\erc}\affiliation{\iow}
\author{W.~H.~Trzaska}\affiliation{\jyv}
\author{M.~Tzanov}\affiliation{\lsu}
\author{M.~Vagins}\affiliation{\uci}
\author{C.~Vilela}\affiliation{\sbu}
\author{Z.~Wang}\affiliation{\tsing}
\author{J.~Wang}\affiliation{\usdm}
\author{M.~Wetstein}\affiliation{\iowa}
\author{M.J.~Wilking}\affiliation{\sbu}
\author{L.~Winslow}\affiliation{\mitnew}
\author{P.~Wittich}\affiliation{\corn}
\author{B.~Wonsak}\affiliation{\ham}
\author{E.~Worcester}\affiliation{\bnl}\affiliation{\sbu}
\author{M.~Wurm}\affiliation{\mainz}
\author{G.~Yang}\affiliation{\sbu}
\author{M.~Yeh}\affiliation{\bnl}
\author{E.D.~Zimmerman}\affiliation{\boul}
\author{S.~Zsoldos}\affiliation{\ucb}\affiliation{\lbnl}
\author{K.~Zuber}\affiliation{\dres}

\newcommand{\jul}{Institut f{\"u}r Kernphysik, Forschungszentrum J{\"u}lich, 52425 J{\"u}lich, Germany
}

\newcommand{\aach}{III. Physikalisches Institut B, RWTH Aachen University, 52062 Aachen, Germany}

\collaboration{Co-Signers from Theia }
\author{T. Akindele}\affiliation{\llnl}
\author{T. Anderson}\affiliation{\psu}
\author{M.~Askins}\affiliation{\lbnl}\affiliation{\ucb}
\author{A. Baldoni}\affiliation{\psu}
\author{A. Barna}\affiliation{\uh}
\author{T. Benson}\affiliation{\Wisconsin}
\author{M. Bergevin	}\affiliation{\llnl}
\author{A. Bernstein	}\affiliation{\llnl}
\author{B. Birrittella	}\affiliation{\Wisconsin}
\author{J. Boissevain	}\affiliation{\penn}
\author{J. Borusinki	}\affiliation{\uh}
\author{D. Cowen	}\affiliation{\psu}
\author{B. Crow	}\affiliation{\uh}
\author{F. Dalnoki-Veress}\affiliation{\miis}
\author{D. Danielson}\affiliation{\chic}
\author{S. Dazeley	}\affiliation{\llnl}
\author{M. Diwan	}\affiliation{\bnl}
\author{A. Druetzler	}\affiliation{\uh}
\author{S. Dye	}\affiliation{\uh}
\author{A. Fienberg	}\affiliation{\psu}
\author{V. Fischer	}\affiliation{\ucd}
\author{K. (Kat) Frankiewicz}\affiliation{\bu}
\author{D. Gooding	}\affiliation{\bu}
\author{C. Graham	}\affiliation{\umich}
\author{C. Grant	}\affiliation{\bu}
\author{J. Griskevich	}\affiliation{\uci}
\author{J. He	}\affiliation{\ucd}
\author{J. Hecla	}\affiliation{\ucb}\affiliation{\llnl}
\author{I. Jovanovic	}\affiliation{\umich}
\author{M. Keenan	}\affiliation{\psu}
\author{P. Keener	}\affiliation{\penn}
\author{P. Kunkle	}\affiliation{\bu}
\author{J. Learned	}\affiliation{\uh}
\author{V. Li	}\affiliation{\llnl}
\author{J. Maricic	}\affiliation{\uh}
\author{P. Marr-Laundrie	}\affiliation{\Wisconsin}
\author{J. Moore	}\affiliation{\psu}
\author{A. Mullen	}\affiliation{\ucb}\affiliation{\llnl}
\author{E. Neights	}\affiliation{\psu}
\author{K. Nishimura	}\affiliation{\uh}
\author{B. O'Meara	}\affiliation{\psu}
\author{K. Ogren	}\affiliation{\umich}
\author{G. D. Orebi Gann	}\affiliation{\ucb}\affiliation{\lbnl}
\author{L. Oxborough	}\affiliation{\Wisconsin}
\author{A. Papatyi	}\affiliation{\pnnl}
\author{B. Paulos	}\affiliation{\Wisconsin}
\author{T. Pershing	}\affiliation{\llnl}
\author{L. Pickard	}\affiliation{\ucd}
\author{L. Sabarots	}\affiliation{\Wisconsin}
\author{V. Shebalin	}\affiliation{\uh}
\author{M. Smy	}\affiliation{\uci}
\author{H. Song	}\affiliation{\bu}
\author{F. Sutanto	}\affiliation{\llnl}
\author{R. Svoboda	}\affiliation{\ucd}
\author{M. Vagins	}\affiliation{\uci}
\author{R. (Rick) Van Berg}\affiliation{\penn}
\author{G. Varner	}\affiliation{\uh}
\author{V. Veeraraghavan}\affiliation{\iowa}
\author{S. Ventura	}\affiliation{\uh}
\author{B. Walsh	}\affiliation{\bnl}
\author{G. Wendel	}\affiliation{\psu}
\author{D. Westphal	}\affiliation{\llnl}
\author{M. Wetstein	}\affiliation{\iowa}
\author{A. Wilhelm	}\affiliation{\umich}
\author{S. Wolcott	}\affiliation{\Wisconsin}
\author{M. Yeh	}\affiliation{\bnl}

\newcommand{\umich}{University of Michigan, Ann Arbor, MI, USA}
\newcommand{\miis}{Middlebury Institute of International Studies at Monterey, Monterey, California 93940, USA}

\newcommand{\usdm}{Department of Physics, South Dakota School of Mines and Technology, Rapid City, SD 57701, USA}
\newcommand{\Wisconsin}{University~of~Wisconsin, Madison, Wisconsin 53706}

\collaboration{Co-Signers from WATCHMAN }

\maketitle

\end{titlepage}

\newpage

\section*{Executive Summary}

\begin{itemize}

\item The High Energy Physics community can benefit from a natural synergy in research activities into next-generation large-scale water and scintillator neutrino detectors, now being studied for remote reactor monitoring, discovery and exclusion applications in cooperative nonproliferation contexts. 
\item Since approximately 2010, US nonproliferation researchers, supported by the National Nuclear Security Administration (NNSA), have been studying a range of possible applications of relatively large (100 ton) to very large (hundreds of kiloton) water and scintillator neutrino detectors.    

\item In parallel, the fundamental physics community has been developing detectors at similar scales and with similar design features for a range of high-priority physics topics, primarily in fundamental neutrino physics. These topics include neutrino oscillation studies at beams and reactors, solar, and geological neutrino measurements,  supernova studies, and others. 
\item Examples of ongoing synergistic work at U.S. national laboratories and universities include prototype gadolinium-doped water and water-based and opaque scintillator test-beds and demonstrators, extensive testing and industry partnerships related to large area fast position-sensitive photomultiplier tubes, and the development of concepts for a possible underground kiloton-scale water-based detector for reactor monitoring and technology demonstrations.
\item Some opportunities for engagement between the two communities include bi-annual Applied Antineutrino Physics conferences, collaboration with U.S. National Laboratories engaging in this research, and occasional NNSA funding opportunities supporting a blend of nonproliferation and basic science R\&D, directed at the U.S. academic community.  

\end{itemize}

\newpage

\section{Introduction}
The High Energy Physics community can benefit from a natural synergy in research activities into next-generation large-scale water and scintillator neutrino detectors, now being studied for remote reactor monitoring, discovery and exclusion applications in cooperative nonproliferation contexts. 
 
Since approximately 2010, US nonproliferation researchers, supported by the National Nuclear Security Administration (NNSA), have been studying a range of possible applications of relatively large (100 ton) to very large (hundreds of kiloton) water and scintillator neutrino detectors.  Primarily due to investments by NNSA, but also with important HEP contributions, a significant R\&D effort has arisen at US National Laboratories, working with academic collaborators to understand and advance the nonproliferation and fundamental science potential of such detectors.  In this white paper, we offer a broad and non-exhaustive summary of areas of actual and potential overlap between HEP fundamental science priorities and NNSA nonproliferation priorities, and suggest ways to continue and formalize the exploitation of the synergies that are already in place. 

In nonproliferation contexts, it is often difficult to reconcile the conflicting goals of non-intrusiveness and robust verification of the absence of illicit nuclear programs. One important area of interest is to confirm either the absence, or conversely the licit operation, of nuclear reactors - the sources of all the world’s plutonium.  Antineutrinos, whose penetrating signature of nuclear origin can provide insight into the operations and characteristics of nuclear reactors, offer a promising path towards remote and non-intrusive, but still robust and persistent verification of reactor operations. These applications have been explored in the last decade, and have resulted in ongoing US and international (primarily UK) collaborative efforts to advance relevant technologies, as well as a deeper understanding of the potential and limitations of those technologies for nonproliferation.  In parallel, the high energy physics community is pursuing similar technical goals that advance fundamental physics, in the areas of MeV-scale solar, atmospheric and geological neutrinos, supernova detection, neutrinoless double beta decay, and the GeV-scale high energy neutrino oscillation and CP violation physics that are the main goals of the US’s flagship Deep Underground Neutrino Experiment (DUNE)~\cite{dune}. 
In this white paper, we provide an incomplete survey of ongoing work in this overlap area, and suggest possible approaches to facilitate cooperation and synergy among the two communities.

\section{Fundamental Physics Relevance }
A range of fundamental physics applications make use of large unsegmented liquid detectors – from the 100 ton to the hundred kiloton scale. Water-based detectors such as Hyper-Kamiokande~\cite{abe2011letter}, scintillator detectors such as JUNO~\cite{juno}, and proposed hybrid or alternative media such as Theia~\cite{Theia} and Liquid-O~\cite{liquido} make use of broadly similar technologies.  Science programs making use of these technologies are far-reaching and already popular in the HEP community. They include such goals as neutrinoless double beta decay, geoneutrino, solar neutrino and supernova physics at the MeV scale, and atmospheric neutrino and beam neutrino oscillation and CP-violation physics at the GeV scale. As presented in the 2019 DUNE Module of Opportunity workshop~\cite{moo}, such detectors are an attractive companion technology to the liquid argon detectors now envisioned for DUNE, with different systematic considerations for accelerator physics goals, and a different and lower energy regime made available by the hydrogenous target, relevant to a complementary set of physics goals. 

\section{Nonproliferation Applications}
The monitoring of nuclear reactor operations, and the discovery or exclusion of unsanctioned plutonium production reactors are central concerns of the global nuclear nonproliferation regime. For convenience we define two standoff domains with their own technological approaches. 
In the mid-field, from 2 kilometers to 20 kilometers,  100 ton to tens of kiloton detectors can accumulate statistics sufficient to identify or exclude the existence of operating reactors, and to track basic features of their operation such as operational status, thermal power levels and fissile content. The interaction channel used is the inverse beta decay (IBD) process, in which a charged-current interaction of an antineutrino with a quasi-free proton generates a positron and a neutron, both of which can be detected in close time coincidence.  While one must be mindful of potentially severe statistical and technology limitations, it may also be possible to extract an independent estimate of the range of the reactor based on a shape analysis of an acquired antineutrino spectrum, and of the direction of the reactor based on the collective displacements between reconstructed positions of the prompt positron emission and the delayed neutron capture in the IBD channel. This set of applications maps well into existing international protocols for monitoring reactors and other nuclear sites, in large part administered by the International Atomic Energy Agency. 
   
In the far-field, from 20-200 kilometers, the small number of detectable IBD interactions in reasonable time periods restricts the applications to the exclusion or discovery of operating reactors, using detectors at the 10 kiloton scale, up to a limit of roughly megaton of target mass.  Discovery and exclusion applications are relevant as confidence-building measures within the framework of existing treaties, and as more robust verification measures called for by potential future treaties such as the Fissile Material Cutoff Treaty~\cite{fmct}  and the Treaty on the Prohibition of Nuclear Weapons~\cite{tpnw}.

\section{Mid-field and Far-field Technology Options for Nonproliferation overlap with Neutrino Physics Experimental Needs}

Currently and for  the foreseeable future, only liquid scintillator and water Cherenkov detectors (and possibly hybrids of these media) appear to have the potential to be practical at the hundreds of ton to hundreds of kiloton scale needed for mid-to-far-field monitoring. Kiloton-scale and larger detectors of both types have been built beginning in the 1970s, with increasing size and performance. The paradigm for reactor antineutrino detection is the kiloton-scale KamLAND liquid scintillator detector, which has operated for two decades, and has detected antineutrinos from reactors across Japan and South Korea, up to hundreds of kilometers away~\cite{KamLAND:2013rgu}.  

Attractions of scintillator detectors include high light output and consequent reduction in the number of sensors needed to capture the photons created by antineutrino interactions, and the ability to thoroughly purify the organic liquid of contaminating backgrounds. Disadvantages are the expense of the medium, in some cases the combustibility or toxicity of the organic materials comprising the scintillator, the need for transport or local manufacture of the scintillator at the deployment site, and the resultant additional complexity these imply for underground deployment and decommissioning of large-scale detectors. Moreover, since the scintillator absorbs light, it is difficult to build detectors with masses larger than a few tens of kilotons of the detector material before light absorption within the medium compromises detection. This constraining characteristic ultimately limits the effective standoff range for scintillator detectors, largely - though not entirely - relegating them to mid-field nonproliferation applications.

Detector masses ranging from tens of tons to tens of kilotons can address some mid-field and far-field applications. The largest currently planned liquid scintillator detector is JUNO~\cite{juno}, comprising 20 kilotons of scintillator immersed in a 35 kiloton water veto, and now being installed in an underground mine in China. JUNO, once operational, will therefore demonstrate a capability relevant for some mid-to-far-field applications. 

For far-field deployments, with the detector located ~20-200 kilometers away from the reactor, 10-1000 kiloton detectors are required to achieve reasonable dwell times to discover or exclude the presence of 50MWt reactors.  At these scales, water-based detectors would offer an attractive alternative to scintillator-based detectors. Water has the advantages of low cost per unit volume, ease of purification on large scales, and fewer adverse effects on health, safety, and the environment than liquid scintillator. The low cost of the raw material competes with the requirement for more high cost photosensors per unit of volume, needed to collect the smaller amount of light generated by particle interactions in water compared to scintillator. 

Super-Kamiokande~\cite{SK} is the largest water Cherenkov detector ever built, with a total mass of water of 50 kilotons.  Hyper-Kamiokande\cite{abe2011letter} is approximately five times larger than Super-Kamiokande, and has been approved for deployment in Japan. Such a detector is on  the scale required for remote exclusion and discovery of reactors at the ~100 kilometer range. The SNO+ experiment\cite{snoibd} also anticipates collection of IBD events from distant reactors, providing further insight into sensitivity of large water-based detectors to the reactor IBD signal.

Pure water detectors provide sub-optimal spectral information about reactor antineutrinos. Two variants on water detectors show promise.  First, one part per thousand quantities of the element gadolinium, a highly effective neutron   capture agent, can be added to water, significantly enhancing sensitivity to the neutron generated in antineutrino interactions. This has been demonstrated in the EGADS 200-ton test-bed\cite{Marti:2019dof} , and the SuperKamiokande experiment has recently added gadolinium  in order to improve sensitivity to antineutrinos of astrophysical origin\cite{SKGd}. The WATCHMAN collaboration has proposed a design\cite{WATCHMAN:2015lcq} for a Gd-H2O detector as a candidate for a demonstration of remote monitoring. 
Such a Water Cherenkov  detector could provide the added potential benefit of direction reconstruction, via antineutrino-electron scattering. Solar Neutrino Electron Scattering has been demonstrated clearly in Super-Kamiokande, but this capability is difficult to achieve and as yet unproven for reactor   antineutrinos. This is expected to be an extremely challenging measurement using a water-based target~\cite{Hellfeld:2015xym}, due to as-yet undemonstrated radio-purity requirements.

Another promising approach is to try to combine the advantages of scintillator and water in a single medium. Water-based scintillators are water detectors in which part per hundred or greater quantities of scintillator are stably mixed, creating both scintillation and Cherenkov light and in principle combining the advantages of both materials and readout modes.   Such detectors may hold promise as a mid-scale solution for reactor antineutrino detection. They have also been explored for close to a decade as one possible target material for the large-scale Theia  detector for a range of fundamental physics experiments~\cite{Theia}.

In addition to  Cherenkov/scintillation light separation, it is also possible to exploit more traditional pulse-shape discrimination (PSD) methods in a suitable formulated WbLS cocktail.  PSD has been used in pure liquid scintillators for decades to distinguish neutrons from gamma-rays with high reliability. Here, the fast and slow decay times of scintillation light arise from the excitation and decay of singlet and triplet states respectively. Neutrons and gamma-rays induce different triplet/singlet ratios in media due to their very different track ionization densities (DE/dx), with the result that pulse time profile differs substantially according to particle type. The use of  scintillation PSD, possibly in combination with  Cherenkov/scintillation separation, should allow for better particle identification for a larger range of particles. Another advantage of incorporating a 'standard' PSD capability into WbLS is that it may be possible to use slower-response light compared to methods relying on Cherenkov/scintillation discrimination.   Recently, researchers at LLNL have developed and demonstrated a  novel majority-water WbLS cocktail, which discriminates between gamma-rays and neutrons using scintillation-based PSD ~\cite{Fordpsd}.

Another early-stage technology of interest is to use opaque scintillating oil (or possibly a water-scintillator combination) instrumented with solid-state photo-detectors coupled to wavelength shifting fibers immersed in the medium. This idea is being explored by the Liquid-O proto-collaboration~\cite{liquido}. The localized nature of the light collection may increase the fiducial-to-total mass ratio of the detector, and may offer significant reductions in the self-shielding and overburden requirements for reactor antineutrino detection through efficient positron identification.

In sum, scintillator detectors have been demonstrated at the kiloton scale, scintillator and modified water detectors appear to hold promise up to the few tens of kiloton scale, suitable for the longer standoff distances in the mid-field, while large gadolinium-doped or scintillator-doped water designs could permit construction of the 100 kiloton and larger detectors that are required for the far-field.

\section{Opportunities for Collaboration }

An excellent existing platform for exploration of the synergies described above is the Applied Antineutrino Physics workshop series\cite{aap}. Since 2004, these conferences have showcased developments in nonproliferation and in particle and nuclear physics that have the potential to influence the other domain.  The conferences have been held in the U.S.  and worldwide, and frequently include updates from representatives of particle physics experiments. The forum is an opportunity to learn the basic working concepts and priorities of the nuclear nonproliferation and disarmament communities, as well as providing nonproliferation experts with insight into cutting edge experimental physics projects worldwide.  

Current examples of ongoing synergistic work at U.S. national laboratories and universities include prototype gadolinium-doped water and water-based and opaque scintillator test-beds and demonstrators, extensive testing and industry partnerships related to large area fast position-sensitive photomultiplier tubes, and the development of concepts for a possible underground kiloton-scale water-based detector for reactor monitoring and technology demonstrations.
The latter collaborative planning and research effort has served as a training ground for physics, nuclear engineering and policy students and postdoctoral fellows, who work side-by-side on technical problems and policy matters related to both nonproliferation and fundamental science.  

NNSA also occasionally issues funding opportunities for academic collaborative efforts focused on nonproliferation and overlapping fundamental science efforts\cite{consortia}.

In the future, proposals that recognize and capitalize on the synergies between the nonproliferation and HEP communities could (and should) be considered for joint and combined funding —shared between the Office of High Energy Physics and NNSA. Pilot investments of this kind have been initiated in the past several years, and the horizon for large-scale neutrino and particle astrophysics experiments suggests that more collaborative opportunities may become available.  

\medskip

Specific areas of possible synergistic research include, but are not limited to: 

\begin{itemize}
\item	Joint technology demonstrations at the 0.1 to 10 kiloton scale;
\item	Further development of water-based scintillator, opaque scintillator, wavelength shifting chemicals added to water, and similar material advances; 
\item	Development of high spatial and temporal resolution photosensors such as the Large Area Picosecond Photodetector (LAPPD)\cite{LAPPDtiming} ; 
\item	Development of affordable instrumentation for large channel counts; 
\item	Efficient and low-cost light collection across an increased wavelength regime extending into toward the IR, such as wavelength shifting plates;
\item	Multi-wavelength-discriminating sensors such as dichroicons\cite{dichroicon}, capable of manufacture and deployment on large scales. 
\item	Development of new or improved scintillator materials, producing more fast light output as well as efficient particle identification (PSD).
\item	Development of new low energy detector types which permit some resolution of neutrino direction, including scaling up from ton scale to kiloton scale.
\item	Development of neutrino detectors capable of geoneutrino observation (\textless 2.8 MeV), low-end reactor neutrino energies, and relic supernova neutrino studies.
\item	Detector development for neutrino spectral measurements extending down to the 1.8 MeV IBD threshold with few percent energy resolution enabling neutrino age measurement via spectrum modulation (neutrino ranging).
\item	Studies of cost effective very large IBD neutrino detectors, capable of long range reactor detection (\textgreater 100 km), nucleon decay  searches reaching to lifetimes in the $10^{35}$ year range, and astrophysical neutrinos. Such detectors have important astrophysical observation capability (starburst galaxies and relic neutrinos), and supernova early warning capability via increased neutron appearance rate detection.
\item	Development of technology for large volume neutrino detectors employing replaceable optical modules without in-liquid wired connections. 

\end{itemize}

\newpage

\bibliography{refs}

\end{document}